\documentclass{emulateapj}
\usepackage{apjfonts}
\usepackage{natbib}
\usepackage{graphicx}
\usepackage{amssymb,amsmath}
\usepackage{xcolor}
\usepackage[export]{adjustbox}
\usepackage[citebordercolor=green]{hyperref}
\shorttitle{Gibbs Sampling}
\usepackage{array}
\newcolumntype{P}[1]{>{\centering\arraybackslash}p{#1}}

\begin{document}

\title{On the Impact of  Random Residual Calibration Error on the Gibbs ILC
Estimates over Large Angular Scales}
\author{Vipin Sudevan\altaffilmark{1}, Rajib Saha\altaffilmark{1}}
\altaffiltext{1}{Physics Department, Indian Institute of Science
Education and Research Bhopal,  Bhopal, M.P, 462023, India.}

\begin{abstract}
Residual error in calibration coefficients corresponding to observed CMB maps is an important 
issue while estimating a pure CMB signal. A component separation method, if these errors in 
 the input foreground contaminated CMB maps are not properly taken into account, may 
lead to bias in the cleaned CMB map and estimated CMB angular power spectrum. But the inability 
to exactly determine the calibration coefficients corresponding to each observed CMB map from 
any CMB experiment makes it very difficult to incorporate their exact and actual values 
in a component separation analysis. Hence the effect of any random and residual calibration error 
on the cleaned CMB map and its angular power spectrum  of a component separation problem can only 
be understood by performing detailed Monte Carlo simulations.  In this paper, we investigate the 
impact of using input foreground contaminated CMB maps with random calibration errors on  
posterior density of cleaned CMB map and theoretical CMB angular power spectrum over large 
angular scales of the sky following the Gibbs ILC method proposed by~\cite{Sudevan:2018qyj}.
By performing detailed Monte Carlo simulations of WMAP and Planck temperature anisotropy observations 
with calibration errors compatible with them we show that the best-fit map corresponding to 
posterior maximum is minimally biased in Gibbs ILC method by a CMB normalization bias and residual
foreground bias. The bias in best-fit CMB angular power spectrum with respect to 
the case where no calibration error is present are $\sim 28 \mu K^2$ and $-4.7 \mu K^2$ respectively 
between $2 \le \ell \le 15$ and $16 \le \ell \le 32$. The calibration error induced error in 
best-fit power spectrum  causes
an overall  $6\%$ increase of the net error when added in 
quadrature with the cosmic variance induced error. 
\end{abstract}

%\maketitle

\section{Introduction}

In the era of precision cosmology, for CMB temperature anisotropy and over large angular scales,
it is no longer the  sensitivity of the detectors but the presence of astrophysical foregrounds and
instrumental systematics which hinder the measurement of a pure CMB signal. An accurate CMB signal 
is essential for better understanding the geometry~\citep{Ade:2015bva}, composition of the 
universe~\citep{Goldstein:2002gf} and  renders stringent constraints on cosmological 
parameters~\citep{Ade:2015xua,2013ApJS..208...19H,Aghanim:2018eyx}. A systematic study of the residual 
systematic  errors on top of the already challenging task of foreground removal from CMB maps is of 
even more importance for a component separation method with many 
planned next generation CMB missions~\citep{2011JCAP...07..025K,2014JCAP...02..006A, Hanany:2019lle,Sutin:2018onu,DiValentino:2016foa,2016SPIE.9904E..0WK}, 
which will be designed to detect the signature of very weak  primordial gravitational waves, an
artifact of the initial inflation of the universe. %The challange is  removing foregrounds and other sources of contaminations
%from the observed CMB maps poses a serious challenge. Apart from the detection of primordial
%gravitational waves, an accurate measurement of CMB if of utmost importance for better
%understanding the geometry, composition of the universe and   renders stringent
%constraints on cosmological parameters.

Apart from the presence of foreground contamination in the observed maps due to emissions
by various astrophysical sources, the presence of residual and random calibration errors poses as a
difficult challenge while estimating a pure signal. Although these residual calibration errors 
may be small in a CMB experiment, their presence implies that it is not possible to obtain exact  
values of calibration coefficients corresponding to observed CMB map of each detector. %Planck mission, for example, has constrained the calibration errors at a 
%level of $\lesssim 1.0\%$ accuracy. 
The Planck consortium, for example,  after using advanced photometric calibration 
techniques like spin-synchronous modulation of the CMB orbital dipole~\citep{Adam:2015rua,Ade:2015ads,Akrami:2018jnw} for 
Low-Frequency instrument (LFI) and by using models of planetary atmospheric emissions and time-variable 
CMB orbital dipole~\citep{Adam:2015vua,Aghanim:2018fcm} for High-Frequency Instrument (HFI) 
maps, has constrained the  calibration uncertainties in the Planck full-sky 
surveys. They provided stringent  limit of  $ 0.024$\% calibration error at  HFI 353 GHz~\cite{Akrami:2018jnw,Aghanim:2018fcm} 
Planck map and $\simeq 1$\% and $\sim$ 1.4\% calibration error for 
HFI 545~\cite{Aghanim:2018fcm} and HFI 857~\cite{Adam:2015vua} Planck maps respectively. Similarly WMAP used the dipole modulation of the CMB signal 
due to the observatory's motion around the Sun~\citep{2011ApJS..192...14J,Hinshaw:2003fc,
2013ApJS..208...20B} as a means to calibrate its maps. 
%They provide a conservative estimate of about 0.5\% absolute calibration error for the
%first-year WMAP data and about 0.2\% calibration uncertainty for nine-year WMAP data. 
But it is impossible to determine the exact value of 
the calibration coefficient corresponding to each individual map. %Hence  performing 
%detailed Monte Carlo simulations, where we simulate foreground contaminated 
%maps which mimics the real-time observed CMB maps, is the only way to understand 
%the impact of using incorrect calibration coefficients during component 
%separation. 
Not accounting for such 
calibration errors in the observed maps during foreground minimization procedure
leads to bias while estimating a cleaned CMB signal~\citep{2010MNRAS.401.1602D}. %~{\color{red} REF} has shown that ILC results can become
%significantly biased even in the presence of smallcalibration errors.
It is therefore natural to ask, what would be the impact of using such improperly calibrated foreground
contaminated CMB maps as inputs to a foreground minimization algorithm
on the final cleaned CMB map and its angular power spectrum. In this article, we focus our study on
Gibbs ILC CMB reconstruction method~\citep{Sudevan:2018qyj}, which possesses various interesting properties as
far as CMB reconstruction by removing foregrounds is concerned as described briefly in the following 
discussions and in other references mentioned therein.

In order to remove the foregrounds from CMB observations performed by various satellite missions 
there exist various (foreground) model dependent and independent methods. Since CMB and different
astrophysical components have different emission laws, a component separation method
can utilize these differences to separate the CMB from foregrounds. %Some of the very popular component
%separation techniques %includes a weiner filtering approach, template fitting, a
%Markov-Chain Monte Carlo based %aaproach, Gibbs Sampling approach, independent component analysis.
An important CMB reconstruction method is the Internal-Linear-Combination (ILC) method
~\citep{Tegmark:1995pn,Tegmark:2003ve,2003ApJS..148...97B,Saha:2005aq}
where in order to obtain a  cleaned CMB signal, it is neither required to explicitly 
model the  individual foreground component's physical morphology 
in the form of templates at some reference frequencies nor in the form of corresponding frequency spectra.
The method is based on the assumption that the frequency spectra of the foregrounds are different from
the frequency spectrum of the CMB, which is assumed to be black-body
in nature~\citep{1994ApJ...420..439M,1996ApJ...473..576F}. 
In the ILC method, a cleaned CMB map is obtained by linearly combining
multi-frequency observed foreground contaminated CMB maps using some amplitude terms knows as weight
factors. These weights follow a constrain that their sum  should be unity and can be
estimated analytically by performing a constrained  minimization of
the variance of the cleaned CMB map.

In recent years the ILC method has been investigated extensively~\citep{Hinshaw:2006ia,Eriksen2004jg,
Saha2011,Saha:2015xya,Sudevan:2016aqu}.~\cite{Sudevan:2017una} proposed a global
ILC method in pixel space by taking into account prior information of
CMB covariance matrix under the assumption that detector noise can be ignored over
the large angular scales of the sky.~\cite{Sudevan:2018qyj} proposed a method to estimate the CMB posterior
density and CMB theoretical angular power spectrum given the observed
data over the large angular scales of the sky in a (foreground) model
independent manner using the ILC method discussed in~\cite{Sudevan:2017una} implemented in harmonic space.
They provided the best fit estimates of both, CMB map and theoretical angular power 
spectrum along with their confidence interval regions and
estimated CMB posterior without any need of explicitly modeling the foreground components.
The theoretical power spectrum results and its error estimates can directly be
integrated to cosmological parameter estimation process.

In Section~\ref{formalism}, we review the basic idea of Gibbs ILC method. In Section~\ref{res_cal_error} 
we  discuss how
the calibration errors affect  the cleaned CMB map. We describe our
Monte Carlo simulations to study the effect of calibration errors  in Section~\ref{Methodology} 
and show the simulation results in Section~\ref{Results}.  In Section~\ref{Conclusion} we
discuss our results  and conclude.

\section{Formalism}
\label{formalism}

In~\cite{Sudevan:2018qyj}, we outlined in detail the formalism of the Gibbs ILC method which was implemented 
in harmonic space at large angular scales on  WMAP and Planck foreground contaminated CMB maps at pixel resolution 
defined by HEALPix~\footnote{Hierarchical Equal Area Iso-Latitude Pixelation of Sphere which was
developed by~\cite{Gorski2005}.} 
pixel resolution parameter $N_{\tt side} = 16$ and at a beam resolution of a Gaussian 
beam with FWHM = $9^{\circ}$. For completeness, in this article we briefly review the method. In the Gibbs ILC approach, 
we estimate the CMB posterior density, 
$P({\bf S},C_{\ell}|{\bf D})$ where {\bf S} is the true CMB signal, $C_{\ell}$ denotes the theoretical CMB angular 
power spectrum and ${\bf D}$  is the given observed CMB data, by drawing samples of {\bf S} 
and $C_{\ell}$ from the distribution using Gibbs sampling technique
~\citep{Larson:2006ds,Eriksen:2004ss,Eriksen:2007mx,2009arXiv0905.3823G,Gibbs1984}. 
In Gibbs sampling at the beginning of any Gibbs iteration $i$, %while estimating the joint density of CMB map, 
%${\bf S}$, and theoretical CMB angular power spectrum, $C_{\ell}$, the CMB posterior density 
%$P({\bf S},C_{\ell}|{\bf D})$ given the observed CMB data ${\bf D}$, 
a  CMB signal ${\bf S}^{i+1}$ is sampled from the 
conditional density of {\bf S}, $P_{1}({\bf S}|{\bf D}, C_{\ell})$, given both observed data ${\bf D}$ and a 
theoretical CMB angular power spectrum $C_{\ell}^{i}$ i.e., 
\begin{equation}
{\bf S}^{i+1} \leftarrow P_{1}({\bf S}|{\bf D}, C_{\ell}^{i})\, .
	\label{P1}
\end{equation}
Using the sampled CMB signal ${\bf S}^{i+1}$, a  theoretical CMB angular power spectrum $C_{\ell}^{i+1}$ 
is sampled from the conditional density of $C_{\ell}$,  $P_{2}(C_{\ell}|{\bf D}, {\bf S})$,  given 
both observed data ${\bf D}$ and a CMB map, ${\bf S}^{i+1}$ i.e.,
\begin{equation}
{C}_{\ell}^{i+1} \leftarrow P_{2}(C_{\ell}|{\bf D}, {\bf S}^{i+1})\, .
	\label{P2}
\end{equation}
At the end of $i^{th}$ iteration, one has a pair of ${\bf S}^{i+1}$ and $C_{\ell}^{i+1}$. 
These two steps are repeated large number of times where at each step $C_{\ell}^{i}$ in Eqn.~\ref{P1} 
is replaced by $C_{\ell}^{i+1}$ of Eqn.~\ref{P2} and similarly ${\bf S}^{i+1}$ of Eqn.~\ref{P2} 
by ${\bf S}^{i+2}$ from new Eqn.~\ref{P1}. Removing 
some initial samples of $\bf S$ and $C_\ell$ (the 
burn-in phase) all other samples in the sequence 
appear as if they are sampled from the joint CMB posterior density $P({\bf S},C_{\ell}|{\bf D})$ 
rather than their individual conditional probability distributions. 

Since we intend to reconstruct the joint CMB posterior density in a foreground model independent 
manner %do not yet poses complete information about the different foregrounds that 
%contaminate the CMB observations, while implementing Gibbs sampling in our Gibbs ILC method,  
%we, instead of sampling a pure CMB signal {\bf S} %, from the conditional density of ${\bf S}$ given both the observed data, 
%${\bf D}$, and a theoretical CMB angular power spectrum, $C_{\ell}$,  
we sample $\bf S$ at each Gibbs iteration by minimizing the foregrounds present in the observed data using 
the global ILC method. 

Let us assume that  we have $n$ number of mean subtracted foreground contaminated full-sky CMB maps ${\bf X}_{i}$, 
at a frequency $\nu_{i}$, with $i = 1,2,...,n$.  Then a CMB estimate $\hat{\bf S}$ of the 
underlying true CMB signal ${\bf S}$ is obtained by linearly combining these $n$ input maps, i.e.,
\begin{equation}
\hat{\bf S} = \sum_{i =1}^{n} w_{i}{\bf X}_{i}\, ,
\label{ilc}
\end{equation}  
where, $w_{i}$ is the weight corresponding to the $i^{\tt th}$ frequency channel. %of a $(1 \times n)$ weight vector. 
To preserve the CMB signal in the cleaned map~\footnote{Ignoring any spectral distortion
CMB temperature anisotropy in thermodynamic temperature unit is independent on frequency.} the weights follow a constrain that the sum of all 
the weights corresponding to $n$ frequency channels should be unity i.e.,
\begin{equation}
\sum_{i=1}^{n} w_{i} = 1\, .
\label{weightconst}
\end{equation}
 Using this condition on weights, 
we perform a constrained minimization of the CMB covariance weighted variance, 
${\sigma}^2 = \hat{\bf S}^{\tt T} {\bf C}^{\tt \dagger} \hat{\bf S}$  
where {\bf C} is the theoretical CMB covariance matrix~\citep{Sudevan:2017una,Sudevan:2018qyj} 
 and  $\dagger$ represents the Moore-Penrose generalized 
inverse~\citep{penrose1955generalized,Penrose1955}, in order to estimate the weights. 
The choice of weights which minimizes $\sigma^{2}$ is obtained by 
following a Lagrange's multiplier approach,
\begin{equation}
	{\bf W} = \frac{\hat {{\bf A}}^{\tt \dagger}{\bf e}}{{\bf e}^{T}{\hat{\bf A}}^{\tt \dagger}{\bf e}}\, ,
\label{gilcmap}
\end{equation}
where, $\hat{ A}_{ij} = {\bf X}^{\tt T}_i{\bf C}^{\tt \dagger}{\bf X}_j$, {\bf W}  is an $(n \times 1)$ weight vector 
and ${\bf e}$ is the $n\times 1$ shape vector of the CMB in thermodynamic temperature units. Typically, 
if the input CMB maps are calibrated correctly across all frequency channels, the shape vector is then an 
$(n \times 1)$ identity column vector, therefore $\sum_{i=1}^{n} w_{i}e_{i} = 1$. The 
cleaned CMB map, $\hat{\bf S}$, estimated using the global ILC method is given by,
\begin{equation}
	\hat{\bf S} = {\bf D W} = {\bf D}\frac{\hat {{\bf A}}^{\tt \dagger}{\bf e}}
	{{\bf e}^{T}{\hat {\bf A}}^{\tt \dagger}{\bf e}}\, ,
	\label{gilcmap1}
\end{equation}
where, {\bf D} is a set of $n$ observed CMB maps $\bigl({\bf X}_1, {\bf X}_2, .., {\bf X}_n \bigr)$. 
Since over large angular scales (e.g., at pixel resolution $N_{\tt side} = 16$ and beam smoothed by a 
Gaussian beam of FWHM = $9^{\circ}$)  
the observed CMB maps have negligible detector noise levels, the global ILC weights
adjust themselves in such a way that they cancel out the correlated foregrounds across frequency 
channels and while doing so they minimize the bad effects of CMB-foreground chance correlations
as well, over large angular scales,  thereby 
providing a cleaned CMB map ${\hat{\bf S}}$ very close to the true CMB signal, ${\bf S}$. Eqn.~\ref{gilcmap} 
shows the relation between global ILC weights, ${\bf W}$, and the CMB shape vector, ${\bf e}$.

%If the CMB signal is not calibrated properly in any of the frequency channel $\nu_{i}$, 
%then in the presence of calibration uncertainties, $\delta_{i}$, corresponding to frequency 
%channel $\nu_{i}$, the shape vector will be modified as $e_{i}^1 = 1 + \delta e_{i}$.
\section{Bias in Presence of Residual Calibration Error}
\label{res_cal_error}
If in a CMB experiment, the observed CMB maps are not calibrated correctly, then in the 
presence of calibration uncertainties $\delta e_i$, corresponding to the CMB map observed 
in the frequency channel $\nu_i$, the elements of CMB shape vector in thermodynamic temperature 
unit will be modified as $e_{i}^\prime = 1 +  \delta e_{i}$, or following a vector notation 
${\bf e}^\prime = {\bf e} + \delta {\bf e}$. If $\delta {\bf e}$ were completely known, the weights 
estimated using the new shape vector ${\bf e}^\prime$, while minimizing the foregrounds 
in these observed maps, will still be subjected to the 
constraint that  $\sum_{i=1}^{n} w_{i}e_{i}^{\prime} = 1$ so that it will not introduce 
any multiplicative bias in the cleaned CMB amplitude. But in any CMB experiment, it is 
not possible to obtain the exact numerical values of calibration 
coefficients corresponding to the observed maps. Hence it is interesting 
to understand what will happen if we use Gibbs ILC method on those 
input maps with some level of calibration uncertainties 
in each map while (incorrectly) assuming the CMB shape vector to be the unit  vector ${\bf e}$
in Eqn.~\ref{gilcmap} (or in Eqn.~\ref{gilcmap1}).

In the presence of calibration error the cleaned map following Eqn.~\ref{ilc} is given by,
\begin{eqnarray}
	\hat{\bf S} = \sum_{i=1}^n\biggl(w_ie^\prime_i{\bf S} + w_ie^\prime_i\sum_{k=1}^{n_f}f^k_i{\bf F}^k_0\biggr) \, ,
	\label{cmap1}
\end{eqnarray}
where ${\bf S}$ and  ${\bf F}^k_0$ respectively represents the true sky CMB signal and foreground  template 
for the foreground component $k$ at some reference frequency and $n_f$ denotes the total 
number of foreground components~\footnote{In Eqn.~\ref{cmap1} we have assumed the detector 
noise is negligible, which is the case for WMAP and Planck observations for temperature 
anisotropy over large angular scales of the sky.}. The factor $f^k_i$ represents the $i^{th}$
element of the $k^{th}$ foreground shape-vector ${\bf f}^k$. Defining, $g^k_i = e^\prime_i f^k_i$ 
we can write Eqn.~\ref{cmap1} following the matrix notation as follows, 
\begin{equation}
	\hat{\bf S} = \Big[{\bf W}^{\tt T} \cdot {\bf e}^{\prime} \Big]\, {\bf S} + \Big[{\bf W}^{T} \cdot   
	\sum_{k=1}^{n_f}{\bf g}^{k} \Big]{\bf F}^{0}_{k} \, . 
\label{calibcmb}
\end{equation} 
Using this equation we  note that in presence of calibration error the foreground shape 
vectors modifies to ${\bf g}^k$ from initial ${\bf f}^k$ without any alteration of total number of foreground 
components or the underlying foreground degrees of freedom. Using Eqn.~\ref{calibcmb}
we can infer about presence of different kinds of bias in presence of calibration error 
as discussed below. 

\subsection{CMB Bias or Normalization Bias}
\label{imp_CMB}
Although, in presence of calibration error ${\bf e}^\prime$ enters in Eqn.~\ref{calibcmb} 
while estimating the weights using Eqn.~\ref{gilcmap}, we assume that 
there are no calibration errors in the observed maps, i.e., we keep ${\bf e}$ as a unit $n\times 1$ 
vector, hence,  $ \Big[{\bf W}^{\tt T} \cdot {\bf e}^{\prime} \Big] \neq 1$ in Eqn.~\ref{calibcmb}. 
This leads to CMB normalization bias in $\hat {\bf S}$. Depending upon whether $ \Big[{\bf W}^{\tt T} \cdot {\bf e}^{\prime} \Big] > 1$ or $< 1$
the CMB map will be biased high or low than the sky CMB signal. We note that, even in presence of 
calibration error, if it so happens $ \Big[{\bf W}^{\tt T} \cdot \delta {\bf e} \Big]  \sim 0$,
then $\Big[{\bf W}^{\tt T} \cdot {\bf e}^{\prime} \Big] \sim \Big[{\bf W}^{\tt T} \cdot {\bf e} \Big]= 1$.
Hence, if random $\delta e_i$ are such that $ \Big[{\bf W}^{\tt T} \cdot \delta {\bf e} \Big]  \sim 0$
the net normalization bias in $\hat {\bf S}$ will be close to zero. A larger deviation of $ \Big[{\bf W}^{\tt T} \cdot \delta {\bf e} \Big] $
from $0$ will lead to greater normalization bias in the cleaned map. 

\begin{table*}
\resizebox{\linewidth}{!}{
\begin{tabular}{*{13}{l}}
%\centering
%\tabletypesize{\small}
%\tablewidth{0pt}
%\tablecolumns{12}
%\tablecaption{Specifications of Input Freqency Maps}
%\startdata
\hline \hline\\
Frequency  &K1   &  30 GHz &  Ka1 & Q  & 44GHz &  V&  70 GHz & W & 100 GHz& 143 GHz &  217 GHz  &  353 GHz  \\  
map        &     &         &      &    &       &   &         &   &        &         &           &            \\  
\hline\\
Calibration  & 0.2  & 0.17 & 0.2  & 0.2 & 0.12  & 0.2 & 0.2  & 0.2  & 0.08 &0.021   & 0.028    &  0.024   \\  
error, $\sigma_c$, (\%)   &      &      &      &     &       &     &      &      &      &        &          &        \\  
\hline \hline\\
\end{tabular}}
{Table 1 - Table contains the level of residual calibration error (in \%) in the CMB maps provided by the WMAP 
and Planck satellite missions.\\ }
\label{table1}
\end{table*}

\subsection{Foreground Bias}
\label{foreg}
In the presence of calibration errors weights are expected 
to be dependent on $\delta{\bf e}$. This may cause weights to deviate from the optimal 
values which would have otherwise removed foregrounds satisfactorily in absence
of the calibration error.  
It is interesting, therefore, to ask {\it how much foreground bias may be caused 
in the cleaned map due to  calibration errors? } 
Following an analysis similar to~\cite{Sudevan:2018qyj,Saha:2015xya}, we obtain,
\begin{equation}
{\bf W} = \frac{\big({\bf I} - {\bf C}_f{\bf C}_f^{\dagger}\big){\bf e}^\prime}
	{{\bf e}^{\prime T}\big({\bf I} - {\bf C}_f{\bf C}_f^{\dagger}\big){\bf e}^\prime}
	\Big[1 + 2(\delta {\bf e})^{\prime T}{\hat {\bf A}}^\dagger {\bf e}^\prime \Big] - 
	\frac{(\delta {\bf e})^{\prime T}{\hat{ \bf A}}^\dagger}{{\bf e}^{\prime T}{\hat {\bf A}}^\dagger{\bf e}^\prime}\, , 
\label{weightproj}
\end{equation}
where $\bf I$ represents the $n\times n$ identity matrix and ${\bf C}_f$ follows, 
\begin{eqnarray}
	{\bf C}_f = \sum_{\ell =2}^{\ell_{max}}\frac{2\ell+1}{C_\ell}{\bf C}^f_\ell \, ,
\end{eqnarray}
as in~\cite{Sudevan:2018qyj} and ${\bf C}^f_\ell$ represents the $n \times n$ empirical foreground 
covariance matrix in multipole space in observed data with calibration error. $C_\ell$ represents the 
theoretical CMB angular power spectrum. In zero order of the small calibration error $\delta {\bf e}$
Eqn.~\ref{weightproj} reduces to
\begin{eqnarray}
	{\bf W} \sim \frac{\big({\bf I} - {\bf C}_f{\bf C}_f^{\dagger}\big){\bf e}^\prime} 
         {{\bf e}^{\prime T}\big({\bf I} - {\bf C}_f{\bf C}_f^{\dagger}\big){\bf e}^\prime}\, .
	 \label{weightproj1}
\end{eqnarray}
We note that  ${\bf C}_f{\bf C}_f^{\dagger}$  
is a projector on the column space of ${\bf C}_f$, $\mathcal{C}({\bf C}_f)$. Now following
~\cite{Sudevan:2018qyj}, if $n > n_f$ then the null space of ${\bf C}_f$ is a non-empty set and 
$\big({\bf I} - {\bf C}_f{\bf C}_f^{\dagger}\big)$ is a projector  on the null space. 
From Eqn.~\ref{weightproj1}, we see that, the weight vector ${\bf W}$  (which is actually 
estimated after incorrectly  assuming CMB shape vector to be a unit vector, $\bf e$ in Eqn.~\ref{gilcmap}) satisfies
\begin{equation}
{\bf W}^{T}{\bf g}_{k} \sim  0 \qquad \forall \qquad k\, ,
\label{noresidual}
\end{equation}  
since ${\bf g}^k$ lies completely inside the column space of ${\bf C}_f$. Since any deviation 
of ${\bf W}^{T}{\bf g}_{k} $ from zero (for any $k$) causes foreground residual in the cleaned map, Eqn.~\ref{noresidual} implies 
that, {\it if the  residual calibration error of the input maps are small, there will be only very small
residual foreground bias in the foreground cleaned CMB map even if the weights are 
estimated assuming no calibration error in the input maps.} %We discuss on this issue further
%in Section~\ref{imp_cmbfg2} through Monte Carlo simulations. 

\begin{figure}[th!]
	\centering
	\includegraphics[scale=0.6]{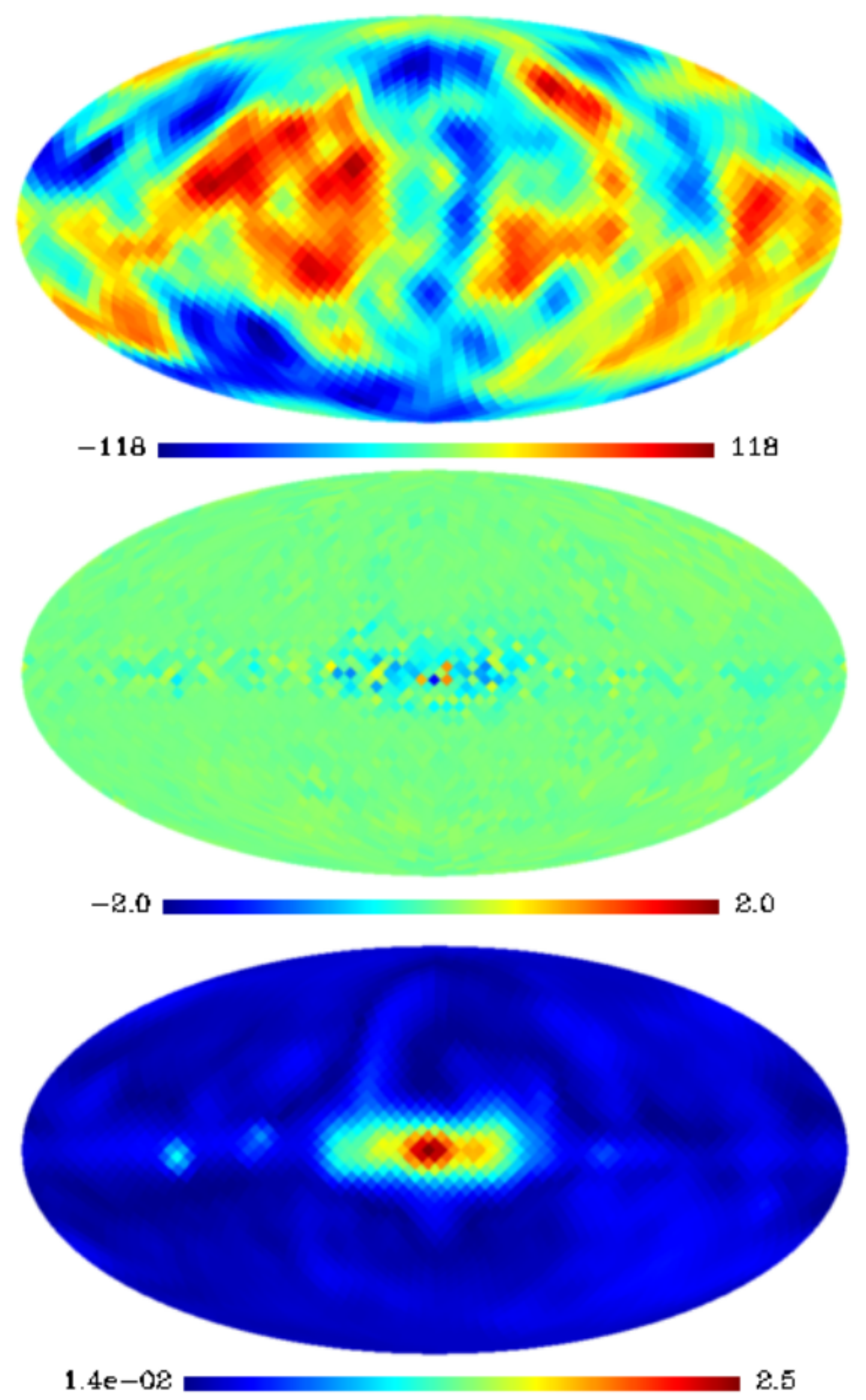}
	\caption{In the top panel, we show the mean of the best-fit cleaned CMB maps obtained from 
		our 1000 Monte Carlo simulations using different set of input simulated foreground contaminated 
		maps with random calibration errors consistent with Planck and WMAP CMB observations. We show 
		the difference between the mean best-fit cleaned CMB map and the best-fit 
		cleaned CMB map obtained from the simulation without any calibration errors in the middle panel.   We see 
		that both the maps agrees well with each other and there is only a minor difference of $2.5\mu K$ in the 
		galactic region. In the bottom panel, the standard deviation of all the 1000 best-fit cleaned CMB maps are shown.}
	\label{error1}
\end{figure}

\begin{figure*}[ht]
        \centering
        \includegraphics[scale=1.5]{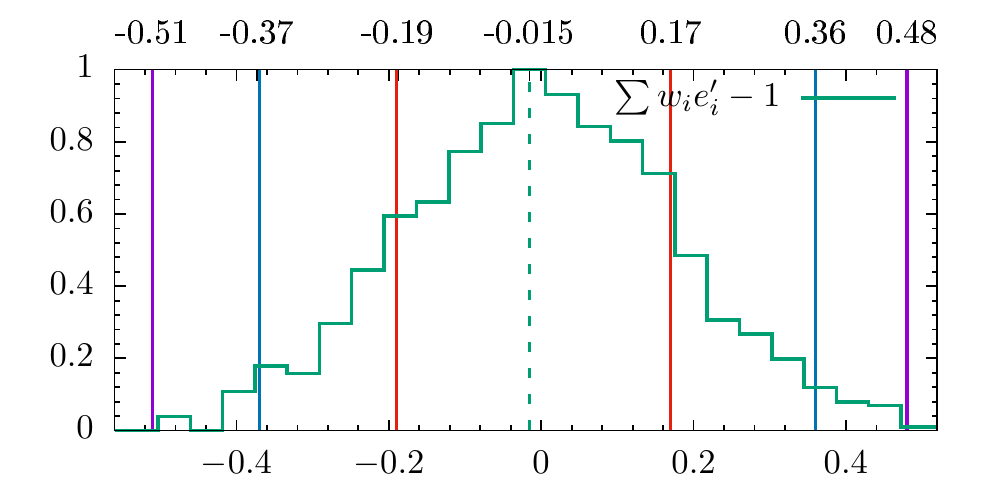}
        \caption{Figure showing the distribution of CMB normalization  bias, $\sum_{i=1}^nw_ie^\prime_i-1$, 
         in percentage level  corresponding to the 
         $1000$ sets of CMB reconstruction using the Gibbs ILC method. We see 
         that the mode of the distribution is centered around $-0.015\%$ which translates into 
         about $\pm 0.00018 \mu$K difference in the temperature at each pixel at 1$\sigma$ level.  }
        \label{error2}
\end{figure*}

\section{Methodology}
\label{Methodology}
The Planck consortium after using advanced photometric calibration techniques like spin-synchronous modulation 
of the CMB orbital dipole for Low-Frequency instrument (LFI) and by using models of planetary 
atmospheric emissions and time-variable CMB orbital dipole for High-Frequency Instrument (HFI) 
maps, has dramatically brought down the calibration uncertainties in the Planck full-sky 
surveys. %They provided an upper limit of about 1\% calibration error till HFI 353 GHz 
%Planck map and 1-1.6\% calibration error for HFI 545 and HFI 857 Planck map. 
However, since  it is impossible to determine the exact value of 
the calibration coefficient corresponding to each individual map, performing 
detailed Monte Carlo simulations, where we simulate foreground and detector noise contaminated 
maps which mimic the real-life observed CMB maps, is the only way to understand 
the impact of using incorrect calibration coefficients during a CMB 
reconstruction method. 

In the current analysis we perform $1000$ different sets of Monte Carlo simulations of entire 
Gibbs ILC procedure  for a comprehensive  study of the impact of using input frequency maps with varying level 
of (residual) calibration errors corresponding to different simulation sets, on the Gibbs ILC results. 
The calibration errors used in these simulations are consistent with the WMAP~\citep{2013ApJS..208...20B, 
2011ApJS..192...14J} and Planck 2018 results~\citep{Akrami:2018jnw,Aghanim:2018fcm}.  
We  mention the calibration error levels in table~\ref{table1} assuming they represent $1\sigma$ 
error levels. 

In these 
Monte Carlo simulations, in each set, we simulated foregrounds and 
detector noise contaminated CMB maps at all WMAP and Planck frequency channels at a pixel 
resolution $N_{side} = 16$ and beam smoothed by a Gaussian beam of FWHM $9^{\circ}$. 
The free-free, synchrotron and thermal dust emissions at different frequency channels are 
obtained at $N_{side} = 256$ and at beam resolution $1^{\circ}$ following the procedure as 
described in~\cite{Sudevan:2016aqu}. These maps are then downgraded to $N_{side} = 16 $ and performed 
an additional smoothing by a Gaussian beam of FWHM = $\sqrt{540^2 - 60^2}$ to bring all 
the foreground maps to $9^{\circ}$  beam smoothing. We generated a CMB temperature map 
using the theoretical CMB power spectrum consistent with cosmological parameters obtained 
by Planck collaboration~\cite{Ade:2015xua} at $N_{side} = 16$ 
and beam smoothing of $9^{\circ}$. We follow the same procedure given in~\cite{Sudevan:2016aqu} 
to generate detector noise maps corresponding to each  input map  at $N_{side} = 16$ 
and $9^{\circ}$ smoothing, the detector noise levels being in 
accordance with the estimate provided by WMAP and Planck science team. 
The final simulated foreground contaminated maps at 
different Planck and WMAP frequencies are obtained by linearly combining the CMB, 
various foregrounds and the detector noise maps.  
 
Once we simulated these input maps, they  are then scaled by calibration factors obtained 
by randomly drawing a unit  mean Gaussian random variable $x_{i}$, where $i$ = 1,..,$n$ (total 
number of maps), with standard deviation equal to the desired amount of calibration 
error mentioned in Table~\ref{table1}. %After obtaining 12 different realization of $x$, each corresponding to same 
%calibration rms error, we scale the input simulated maps as (1.0 + $x$). 
This generates a given set of input maps with the randomly chosen calibration errors.
For the purpose of Monte Carlo simulations we  simulate a total of $1000$
different sets of input maps with random calibration errors. 
% For each set, the calibration error goes as integer multiples of 0.001 (0.1\% 
%calibration error), i.e., in a set which contains simulated maps with calibration 
%error of 0.2\% implies that each map in that set is scaled by a number lying in between 
%$1~\pm~0.002$ with $68.27\%$ probability. %We have shown the  calibration coefficients corresponding to each simulated 
%foreground and noise contaminated CMB map for each set of Monte Carlo simulations in 
%the Table~\ref{table1}. 

After simulating the foreground contaminated maps with different levels of calibration 
error for different sets, we use Gibbs ILC algorithm to minimize the foregrounds. 
While implementing Gibbs ILC code, we assumed that all the input simulated maps are 
calibrated  equally i.e., the shape vector is a unit vector, ${\bf e}$. In the current implementation 
of Gibbs ILC procedure, each simulation consists of $10$ chains each with randomly chosen initial 
points and $5000$  Gibbs steps. We reject $50$ samples (each from cleaned CMB maps and sampled CMB theoretical
angular power spectra) corresponding to  initial burn-in 
phase. This results in a total of $49500$ samples from each simulation.

\section{Results}
\label{Results}
In this section, we discuss the results obtained after performing detailed Monte 
Carlo simulations of  CMB reconstruction using $1000$ different sets of input maps with random calibration 
errors consistent with WMAP and Planck observations. While implementing the Gibbs ILC method, during the foreground minimization 
using the global ILC method we do not  take into account of the presence of calibration errors 
in the map. We follow the same procedure as outlined in~\cite{Sudevan:2018qyj} for calculating the 
CMB posterior, best-fit CMB map and best-fit theoretical CMB angular power spectrum from the 
cleaned CMB map and theoretical angular power spectrum samples generated.

\subsection{Cleaned Maps}
Using $49500$ sampled maps from a given set of simulations we estimate the best-fit 
CMB map corresponding to the maximum likely pixel values for each pixel. Using  
best-fit CMB maps from all $1000$ simulation sets we estimate a simple mean map.
 In Fig.~\ref{error1}, we show the mean best-fit cleaned CMB map in the top
panel. This map agrees very well with the best-fit CMB map when the simulation
involved no calibration error in input maps. This is shown in the middle panel of 
Fig.~\ref{error1}. Minor difference of $\sim \pm 2 \mu K$ is seen only in the 
central galactic plane. The bottom panel of this figure shows the standard deviation 
map computed from $1000$ best-fit CMB maps. From the standard deviation map 
we see that maximum error occurs in the central galactic  region of magnitude
$\sim 2.5\mu K$. Using these results we therefore conclude that even in the presence 
of realistic levels of residual calibration error in the input frequency maps the 
Gibbs ILC method performs very well in CMB reconstruction.  The resulting residual 
foreground contamination due to any unaccounted for calibration error is limited to $\sim \pm 6 \mu K$
with a $3\sigma$ confidence level only along the very central region of the galactic
plane.  

\begin{figure}[th!]
        \centering
        \includegraphics[scale=0.6]{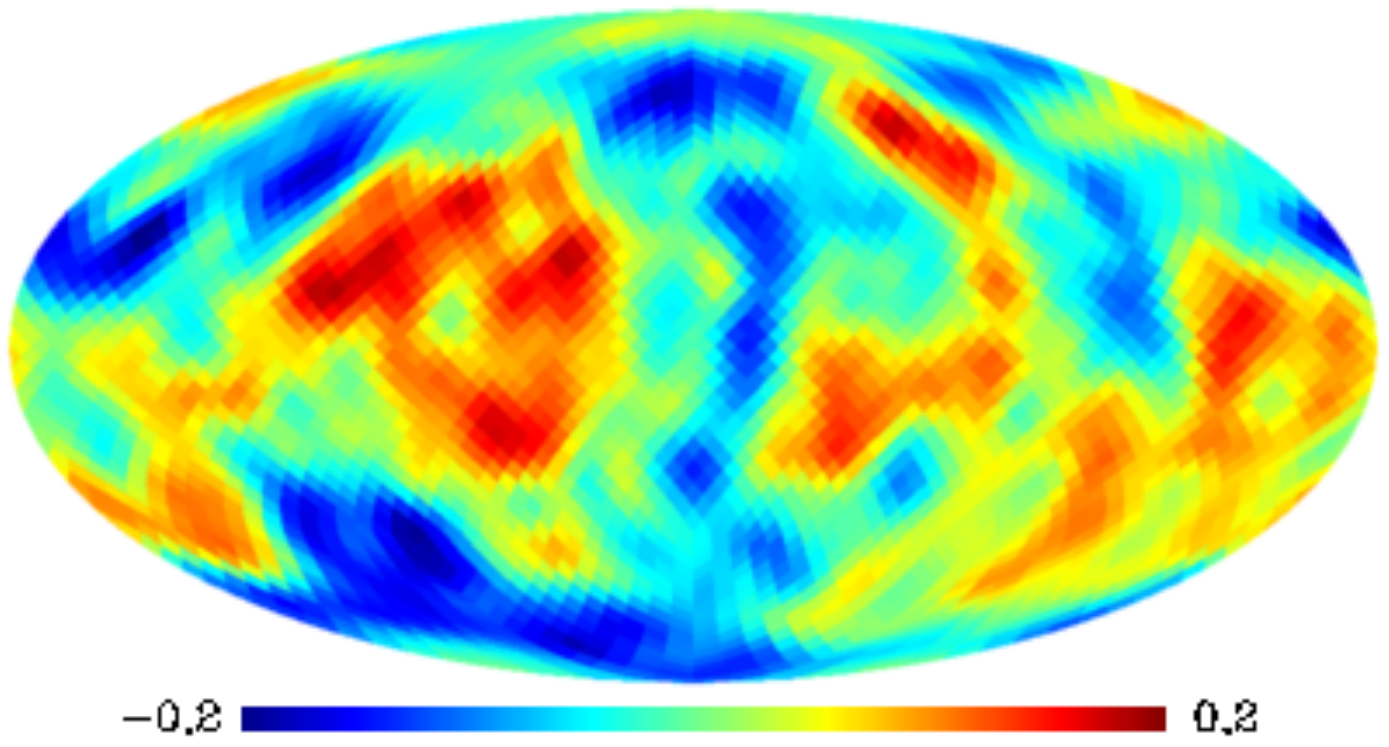}
        \caption{Figure showing the normalization bias in the CMB map  corresponding to the 1$\sigma$ level 
        of $\sum w_{i}e_{i}^{1}-1$. We see that at 1$\sigma$ of normalization bias the maximum change in 
        pixel temperature is of the order of $\pm 0.2 \mu$K. }
        \label{error21}
\end{figure}

\begin{figure*}[th!]
        \centering
        \includegraphics[scale=0.7]{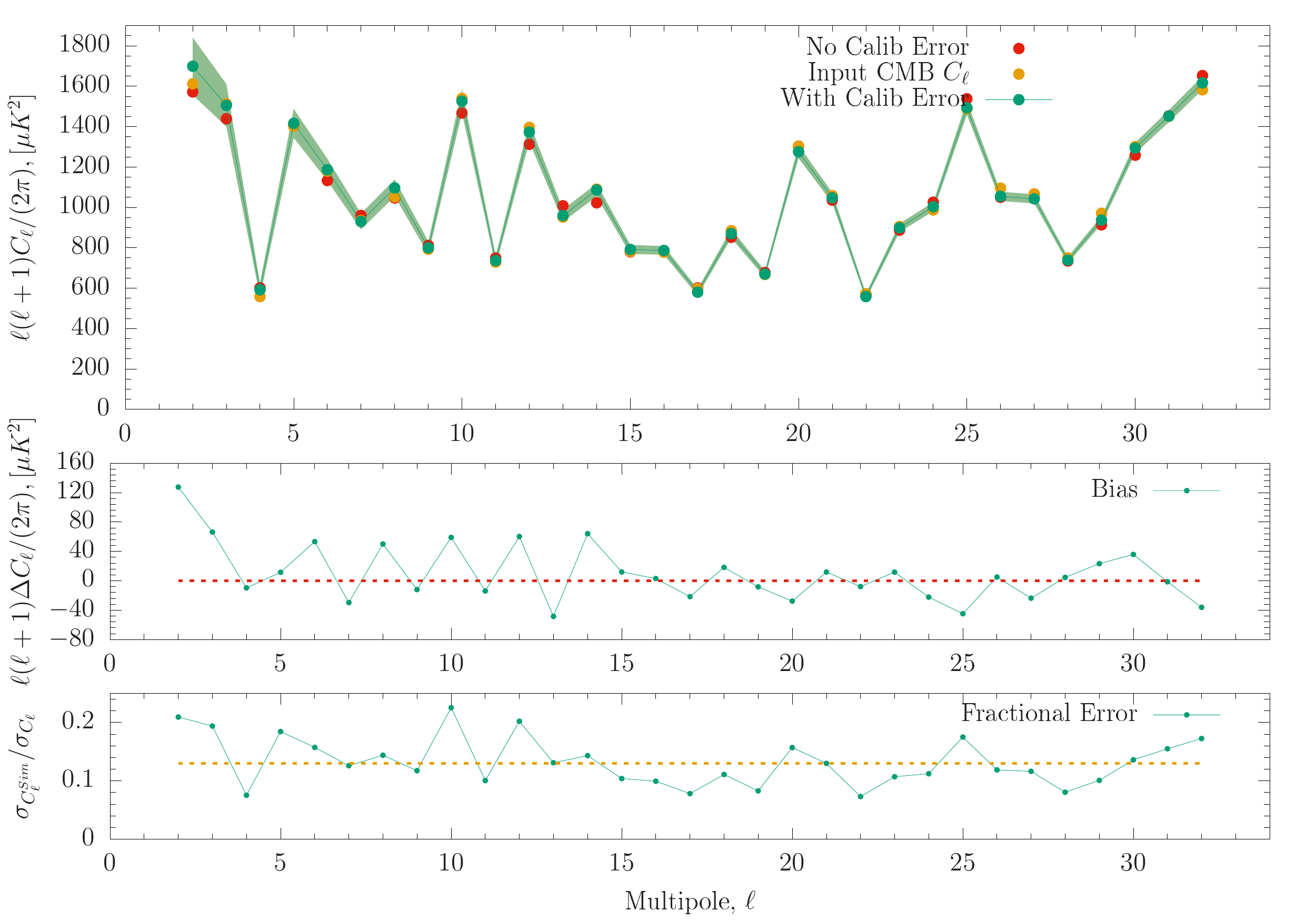}
        \caption{In the top panel, we show the mean best-fit estimated from all the 
        1000 Monte Carlo simulations using green line. The best-fit theoretical angular power spectrum 
        obtained from simulation with no calibration error is shown using red points and the 
        input CMB angular power spectrum used in all the simulations is shown using yellow points. 
        The 1$\sigma$ standard deviation region corresponding to the best-fits 
        from the Monte Carlo simulations is shown as a light green shaded region around the mean
         best-fit angular power spectrum. We see from 
        this plot that the mean best-fit agrees well with both the  input CMB and best-fit (from no 
        calibration errors simulation) angular power spectrum. Both the  input CMB angular power 
        spectrum and best-fit angular power spectrum  
        lie inside the 1$\sigma$ region of the mean best-fit. In the middle panel 
        we show the difference of mean  best-fit angular power spectrum where calibration error 
        was included in the simulations and best fit spectrum without any calibration error. The bottom panel  
         compares the standard deviations of the best-fit spectra with calibration error with the cosmic variance.
        The calibration induced  errors appear to be approximately uniformly distributed  over multipoles
        with respect to the cosmic variance induced errors. Yellow line represents the mean fractional error of $13\%$
        between the entire multipole range.   
        }   
        \label{error3}
\end{figure*}

Apart from the small level of residual foreground bias  due to  residual calibration error 
in input frequency maps as discussed above the cleaned CMB maps contain a CMB normalization bias 
for the same reason (e.g., see Section~\ref{foreg} and~\ref{imp_CMB}). The normalization 
bias arises since in case of input frequency maps with calibration errors weights 
should have satisfied ${\bf W}^T{\bf e}^\prime = 1$ to preserve the amplitude of the CMB 
component in the cleaned map, whereas, they satisfy  ${\bf W}^T{\bf e} = 1$, where 
${\bf e}^\prime = {\bf e} + \delta {\bf e}$ and $\delta {\bf e}$ represents the unknown
residual calibration error. A measure of the normalization bias in any given experiment 
with a given set of unknown residual calibration error is then ${\bf W}^T{\bf e}^\prime - 1$. 
We show the distribution of CMB normalization  bias in percentage level using $1000$ sets of CMB reconstruction 
using the Gibbs ILC method  in Fig.~\ref{error2}. As seen from this figure the normalization bias
is only $0.17\%$ at $1 \sigma$ level in a CMB reconstruction method using Gibbs ILC method.    
In Fig.~\ref{error21} we show the normalization bias map for the chosen  input CMB map 
of this work corresponding to $1\sigma$ value of $\sum w_{i}e_{i}^{1}-1$. For the calibration 
error levels of WMAP and Planck the normalization bias 
is less than $0.2 \mu K$ in magnitude.

\subsection{Best-fit CMB angular power spectrum} 
Since  exact values of calibration coefficients are unknown  in any 
CMB experiment estimated values of  best-fit CMB angular power spectrum 
would be different from the actual ones that would have been estimated 
in the hypothetical case when the calibration coefficients would be 
exactly known. Such differences may result in  a bias in the estimated best-fit 
CMB angular power spectrum apart from  causing larger errors in the later  due 
to random nature of the calibration uncertainties.   Using $1000$ Monte Carlo
simulations of CMB posterior estimations using the Gibbs ILC 
method in this section we assess this bias and error in the best-fit CMB 
angular power spectrum in presence of calibration uncertainties. 

In order to  understand any possible bias in the best-fit CMB power spectrum in Gibbs 
ILC method due to residual calibration error we show in top panel of  Fig.~\ref{error3} the mean best-fit CMB angular power 
spectrum obtained from $1000$ simulations (in green) with calibration errors along 
with the best-fit angular power spectrum when the input frequency maps contained no
calibration error (in red). The green filled region around the mean best-fit angular power 
spectrum represents the $1\sigma$ error levels at different multipoles 
obtained from $1000$ best-fit  angular power spectra. The mean best-fit with calibration error 
matches very well with the zero calibration error case.  In the middle panel of this 
figure we show any bias in the best-fit angular power spectrum with respect to the 
zero calibration error case by by plotting the difference of mean best-fit CMB 
angular power spectrum (with calibration error) and best-fit spectrum without any calibration 
error. Visually the differences take more positive values between $\ell = 2$ and $\ell = 15$,
whereas, differences take somewhat more negative values between $\ell = 16$ to $\ell = 32$.
The average bias in the first multipole range is as small as $~\sim 28 \mu K^2$. Average 
bias between $\ell = 16$ to $\ell = 32$ is just $-4.7 \mu K^2$.   
%From this panel we see that the fractional bias on the best-fit CMB angular power 
%spectrum in presence of calibration error is less than $10\%$ for $\ell =2$. For 
%higher multipoles the fractional bias decreases and becomes just $2\%$ at $\ell = 32$.   
In the bottom panel of Fig.~\ref{error3} we show the calibration uncertainty induced 
error in the best-fit CMB angular power spectrum by plotting the ratio of standard 
deviations ($\sigma_{C^{Sim}_\ell}$) of the best-fit spectra in presence of calibration error and  cosmic
variance induced error ($\sigma_{C_{\ell}}$).  The standard deviation from simulations
varies between $8\%$ (at $\ell = 4$)  and a maximum of $22\%$ at $\ell = 10$ of cosmic variance 
induced error. Calibration induced mean fractional error between $\ell =2$ and $\ell = 32$ is 
just $13\%$.  The calibration error induced error is expected to add in quadrature 
with the cosmic variance induced error. Considering this mean level of fractional error 
net error becomes  $\sqrt{1+0.13} = 1.06$
times of the cosmic variance induced error. This  causes 
a $6\%$ error increase from the cosmic variance prediction between multipole range $2 \le \ell \le 32$.    

Summarizing the simulation results of this section, we conclude that even in the presence 
of calibration uncertainties in the input foreground and detector noise contaminated CMB maps, our Gibbs 
ILC method produces a best-fit cleaned CMB map which has very minor level of residual foreground contamination 
bias and an almost negligible CMB normalization bias. For the best-fit power spectrum  calibration induced  
bias and  error both remain small.

\section{Conclusions \& Discussions}
\label{Conclusion}
The level of calibration uncertainties present in the observed CMB maps have been drastically 
reduced by following advanced photometric calibration techniques like spin-synchronous 
modulation of the CMB orbital dipole~\cite{Adam:2015rua,Ade:2015ads,Akrami:2018jnw}, by using models of 
planetary atmospheric emissions~\cite{Adam:2015vua,Aghanim:2018fcm} 
etc. But the presence of the residual calibration error (however small it be) in the 
observed CMB maps may pose a difficult challenge while estimating a pure CMB signal. 
In the current article, we study the impact of random calibration errors on the 
CMB map and its angular power spectrum  that are obtained using Gibbs ILC method. Since it is impossible to obtain 
the exact value of calibration uncertainties corresponding to each individual observed 
CMB maps, we therefore, perform detailed Monte Carlo simulations of Gibbs ILC method 
with realistic residual calibration errors compatible to WMAP and Planck observations, after simulating 
realistic foreground and detector noise contaminated CMB maps over large 
angular scales of the sky.   

Using analytical results we show in Section~\ref{res_cal_error} that residual errors in calibration coefficients 
lead to two distinct type of bias in Gibbs ILC method which is implemented 
over large angular scales of the sky. The first kind of bias is called the CMB 
normalization bias which arises since the empirical weights satisfy ${\bf W}^T {\bf e} = 1$ 
instead of ${\bf W}^T{\bf e}^\prime =1$, where ${\bf e}^\prime$ represents the true CMB 
shape vector in presence of calibration error. The second type of bias is due to residual
foreground contamination in the cleaned maps. By estimating best-fit cleaned CMB maps 
corresponding to the maximum of the posterior density by detailed Monte Carlo simulations, 
in Sec.~\ref{Results},  the normalization bias is merely $0.17\%$ at a confidence level 
of $1 \sigma$. The residual foreground bias is small as well. Our Monte Carlo 
simulations show that the residual calibration  error tend to maximally bias 
only the galactic central region, with a magnitude of $\sim 2 \mu K$ at $1 \sigma$ confidence
level. Between $2 \le \ell \le 15$ (mean) bias in the best-fit CMB angular power spectrum when calibration errors 
are present in the input maps in Gibbs ILC  method is $\sim 28 \mu K^2$ with respect to the 
ideal case of zero calibration error. This  bias decreases with increase in 
multipoles, and is just $-4.7 \mu K^2$ between  $16 \le \ell \le  32$. The calibration error widens the error intervals
on the best-fit CMB angular spectrum. The average increase of net error level between  $2 \le \ell \le 32$ 
is  $\sim 6\%$ over the cosmic variance induced error.

Based upon our Monte Carlo simulations we conclude that for an analysis over large angular scales 
of the sky, even if we use the maps with 
realistic (residual) calibration errors without accounting for the same in the Gibbs 
ILC algorithm (by modifying the CMB shape vector) leads to very minor level of bias in the 
best-fit cleaned CMB map. The bias and error in the best-fit CMB angular power spectrum 
are both small, however, they may  not completely negligible. It would be important 
to incorporate  such bias and error in the angular power spectrum in cosmological parameter 
estimation and investigate  their role in cosmological parameter estimation.

Finally, we mention an interesting advantage of the Gibbs ILC method on the issue of 
impact of residual calibration errors on the CMB reconstruction.  
Since our method does not require to model the frequency spectrum nor any templates 
for foreground components to reconstruct the CMB  products to a good accuracy our foreground 
removal is independent on the calibration error, as long as the later is small (e.g., 
Section~\ref{foreg}).   

This  work is based on observations obtained with Planck (http://www.esa.int/Planck)
and WMAP (https://map.gsfc.nasa.gov/). Planck was 
an ESA science mission with instruments and contributions directly funded by ESA Member States, 
NASA, and Canada.   
We acknowledge use of Planck Legacy Archive (PLA) and the Legacy Archive for Microwave Background 
Data Analysis (LAMBDA). LAMBDA is a part of the High Energy Astrophysics Science Archive Center (HEASARC). 
HEASARC/LAMBDA is supported by the Astrophysics Science Division at the NASA Goddard Space Flight
Center. We  use publicly available HEALPix~\cite{Gorski2005} package  
(http://healpix.sourceforge.net) for the analysis of this work.

%\bibliography{ms}  

\begin{thebibliography}{}
\expandafter\ifx\csname natexlab\endcsname\relax\def\natexlab#1{#1}\fi

\bibitem[{Adam {et~al.}(2016{\natexlab{a}})}]{Adam:2015rua}
Adam, R., {et~al.} 2016{\natexlab{a}}, Astron. Astrophys., 594, A1

\bibitem[{Adam {et~al.}(2016{\natexlab{b}})}]{Adam:2015vua}
---. 2016{\natexlab{b}}, Astron. Astrophys., 594, A8

\bibitem[{Ade {et~al.}(2016{\natexlab{a}})}]{Ade:2015bva}
Ade, P. A.~R., {et~al.} 2016{\natexlab{a}}, Astron. Astrophys., 594, A18

\bibitem[{Ade {et~al.}(2016{\natexlab{b}})}]{Ade:2015ads}
---. 2016{\natexlab{b}}, Astron. Astrophys., 594, A5

\bibitem[{Ade {et~al.}(2016{\natexlab{c}})}]{Ade:2015xua}
---. 2016{\natexlab{c}}, Astron. Astrophys., 594, A13

\bibitem[{Aghanim {et~al.}(2018{\natexlab{a}})}]{Aghanim:2018fcm}
Aghanim, N., {et~al.} 2018{\natexlab{a}}, arXiv:1807.06207

\bibitem[{Aghanim {et~al.}(2018{\natexlab{b}})}]{Aghanim:2018eyx}
---. 2018{\natexlab{b}}, arXiv:1807.06209

\bibitem[{Akrami {et~al.}(2018)}]{Akrami:2018jnw}
Akrami, Y., {et~al.} 2018, arXiv:1807.06206

\bibitem[{{Andr{\'e}} {et~al.}(2014){Andr{\'e}}, {Baccigalupi}, {Banday},
  {Barbosa}, {Barreiro}, {Bartlett}, {Bartolo}, {Battistelli}, {Battye},
  {Bendo}, {Beno{\^\i}t}, {Bernard}, {Bersanelli}, {B{\'e}thermin},
  {Bielewicz}, {Bonaldi}, {Bouchet}, {Boulanger}, {Brand}, {Bucher},
  {Burigana}, {Cai}, {Camus}, {Casas}, {Casasola}, {Castex}, {Challinor},
  {Chluba}, {Chon}, {Colafrancesco}, {Comis}, {Cuttaia}, {D'Alessandro}, {Da
  Silva}, {Davis}, {de Avillez}, {de Bernardis}, {de Petris}, {de Rosa}, {de
  Zotti}, {Delabrouille}, {D{\'e}sert}, {Dickinson}, {Diego}, {Dunkley},
  {En{\ss}lin}, {Errard}, {Falgarone}, {Ferreira}, {Ferri{\`e}re}, {Finelli},
  {Fletcher}, {Fosalba}, {Fuller}, {Galli}, {Ganga}, {Garc{\'\i}a-Bellido},
  {Ghribi}, {Giard}, {Giraud-H{\'e}raud}, {Gonzalez-Nuevo}, {Grainge},
  {Gruppuso}, {Hall}, {Hamilton}, {Haverkorn}, {Hernandez-Monteagudo},
  {Herranz}, {Jackson}, {Jaffe}, {Khatri}, {Kunz}, {Lamagna}, {Lattanzi},
  {Leahy}, {Lesgourgues}, {Liguori}, {Liuzzo}, {Lopez-Caniego}, {Macias-Perez},
  {Maffei}, {Maino}, {Mangilli}, {Martinez-Gonzalez}, {Martins}, {Masi},
  {Massardi}, {Matarrese}, {Melchiorri}, {Melin}, {Mennella}, {Mignano},
  {Miville-Desch{\^e}nes}, {Monfardini}, {Murphy}, {Naselsky}, {Nati},
  {Natoli}, {Negrello}, {Noviello}, {O'Sullivan}, {Paci}, {Pagano}, {Paladino},
  {Palanque-Delabrouille}, {Paoletti}, {Peiris}, {Perrotta}, {Piacentini},
  {Piat}, {Piccirillo}, {Pisano}, {Polenta}, {Pollo}, {Ponthieu},
  {Remazeilles}, {Ricciardi}, {Roman}, {Rosset}, {Rubino-Martin}, {Salatino},
  {Schillaci}, {Shellard}, {Silk}, {Starobinsky}, {Stompor}, {Sunyaev},
  {Tartari}, {Terenzi}, {Toffolatti}, {Tomasi}, {Trappe}, {Tristram},
  {Trombetti}, {Tucci}, {Van de Weijgaert}, {Van Tent}, {Verde}, {Vielva},
  {Wand elt}, {Watson}, \& {Withington}}]{2014JCAP...02..006A}
{Andr{\'e}}, P., {Baccigalupi}, C., {Banday}, A., {et~al.} 2014, Jour. Cosmol.
  Astro. Phys., 2014, 006

\bibitem[{{Bennett} {et~al.}(2003)}]{2003ApJS..148...97B}
{Bennett}, C.~L., {et~al.} 2003, "Astrophys. J. Suppl.", 148, 97

\bibitem[{{Bennett} {et~al.}(2013){Bennett}, {Larson}, {Weiland}, {Jarosik},
  {Hinshaw}, {Odegard}, {Smith}, {Hill}, {Gold}, {Halpern}, {Komatsu}, {Nolta},
  {Page}, {Spergel}, {Wollack}, {Dunkley}, {Kogut}, {Limon}, {Meyer}, {Tucker},
  \& {Wright}}]{2013ApJS..208...20B}
{Bennett}, C.~L., {Larson}, D., {Weiland}, J.~L., {et~al.} 2013, \apjs, 208, 20

\bibitem[{Di~Valentino {et~al.}(2018)}]{DiValentino:2016foa}
Di~Valentino, E., {et~al.} 2018, JCAP, 1804, 017

\bibitem[{{Dick} {et~al.}(2010){Dick}, {Remazeilles}, \&
  {Delabrouille}}]{2010MNRAS.401.1602D}
{Dick}, J., {Remazeilles}, M., \& {Delabrouille}, J. 2010, \mnras, 401, 1602

\bibitem[{Eriksen {et~al.}(2008)Eriksen, Jewell, Dickinson, Banday, Gorski, \&
  Lawrence}]{Eriksen:2007mx}
Eriksen, H.~K., Jewell, J.~B., Dickinson, C., {et~al.} 2008, Astrophys. J.,
  676, 10

\bibitem[{{Eriksen} {et~al.}(2004)}]{Eriksen2004jg}
{Eriksen}, H.~K., {et~al.} 2004, Astrophys. J., 612, 633

\bibitem[{Eriksen {et~al.}(2004)}]{Eriksen:2004ss}
Eriksen, H.~K., {et~al.} 2004, Astrophys. J. Suppl., 155, 227

\bibitem[{{Fixsen} {et~al.}(1996)}]{1996ApJ...473..576F}
{Fixsen}, D.~J., {et~al.} 1996, "Astrophys. J.", 473, 576

\bibitem[{Geman \& Geman(1984)}]{Gibbs1984}
Geman, S., \& Geman, D. 1984, IEEE Trans. Pattern Anal. Mach. Intell., 6, 721

\bibitem[{Goldstein {et~al.}(2003)}]{Goldstein:2002gf}
Goldstein, J.~H., {et~al.} 2003, Astrophys. J., 599, 773

\bibitem[{{G{\'o}rski} {et~al.}(2005){G{\'o}rski}, {Hivon}, {Banday}, {Wand
  elt}, {Hansen}, {Reinecke}, \& {Bartelmann}}]{Gorski2005}
{G{\'o}rski}, K.~M., {Hivon}, E., {Banday}, A.~J., {et~al.} 2005, \apj, 622,
  759

\bibitem[{{Groeneboom}(2009)}]{2009arXiv0905.3823G}
{Groeneboom}, N.~E. 2009, ArXiv e-prints, arXiv:0905.3823

\bibitem[{Hanany {et~al.}(2019)}]{Hanany:2019lle}
Hanany, S., {et~al.} 2019, arXiv:1902.10541

\bibitem[{Hinshaw {et~al.}(2003)}]{Hinshaw:2003fc}
Hinshaw, G., {et~al.} 2003, Astrophys. J. Suppl., 148, 63

\bibitem[{Hinshaw {et~al.}(2007)}]{Hinshaw:2006ia}
---. 2007, Astrophys. J. Suppl., 170, 288

\bibitem[{{Hinshaw} {et~al.}(2013)}]{2013ApJS..208...19H}
{Hinshaw}, G., {et~al.} 2013, Astrophys. J. Suppl., 208, 19

\bibitem[{{Jarosik} {et~al.}(2011){Jarosik}, {Bennett}, {Dunkley}, {Gold},
  {Greason}, {Halpern}, {Hill}, {Hinshaw}, {Kogut}, {Komatsu}, {Larson},
  {Limon}, {Meyer}, {Nolta}, {Odegard}, {Page}, {Smith}, {Spergel}, {Tucker},
  {Weiland }, {Wollack}, \& {Wright}}]{2011ApJS..192...14J}
{Jarosik}, N., {Bennett}, C.~L., {Dunkley}, J., {et~al.} 2011, \apjs, 192, 14

\bibitem[{{Kogut} {et~al.}(2016){Kogut}, {Chluba}, {Fixsen}, {Meyer}, \&
  {Spergel}}]{2016SPIE.9904E..0WK}
{Kogut}, A., {Chluba}, J., {Fixsen}, D.~J., {Meyer}, S., \& {Spergel}, D. 2016,
  Society of Photo-Optical Instrumentation Engineers (SPIE) Conference Series,
  Vol. 9904, {The Primordial Inflation Explorer (PIXIE)}, 99040W

\bibitem[{{Kogut} {et~al.}(2011)}]{2011JCAP...07..025K}
{Kogut}, A., {et~al.} 2011, Journal of Cosmology and Astro-Particle Physics,
  2011, 025

\bibitem[{Larson {et~al.}(2007)}]{Larson:2006ds}
Larson, D.~L., {et~al.} 2007, Astrophys. J., 656, 653

\bibitem[{{Mather} {et~al.}(1994)}]{1994ApJ...420..439M}
{Mather}, J.~C., {et~al.} 1994, "Astrophys. J.", 420, 439

\bibitem[{Penrose(1955{\natexlab{a}})}]{penrose1955generalized}
Penrose, R. 1955{\natexlab{a}}, in , 406--413

\bibitem[{Penrose(1955{\natexlab{b}})}]{Penrose1955}
Penrose, R. 1955{\natexlab{b}}, Mathematical Proceedings of the Cambridge
  Philosophical Society, 51, 406

\bibitem[{{Saha}(2011)}]{Saha2011}
{Saha}, R. 2011, Astrophys. J. Lett., 739, L56

\bibitem[{Saha \& Aluri(2016)}]{Saha:2015xya}
Saha, R., \& Aluri, P.~K. 2016, Astrophys. J., 829, 113

\bibitem[{Saha {et~al.}(2006)Saha, Jain, \& Souradeep}]{Saha:2005aq}
Saha, R., Jain, P., \& Souradeep, T. 2006, Astrophys. J., 645, L89

\bibitem[{Sudevan {et~al.}(2017)Sudevan, Aluri, Yadav, Saha, \&
  Souradeep}]{Sudevan:2016aqu}
Sudevan, V., Aluri, P.~K., Yadav, S.~K., Saha, R., \& Souradeep, T. 2017,
  Astrophys. J., 842, 62

\bibitem[{Sudevan \& Saha(2018{\natexlab{a}})}]{Sudevan:2017una}
Sudevan, V., \& Saha, R. 2018{\natexlab{a}}, Astrophys. J., 867, 74

\bibitem[{Sudevan \& Saha(2018{\natexlab{b}})}]{Sudevan:2018qyj}
---. 2018{\natexlab{b}}, arXiv:1810.08872

\bibitem[{Sutin {et~al.}(2018)}]{Sutin:2018onu}
Sutin, B.~M., {et~al.} 2018, Proc. SPIE Int. Soc. Opt. Eng., 10698, 106984F

\bibitem[{Tegmark \& Efstathiou(1996)}]{Tegmark:1995pn}
Tegmark, M., \& Efstathiou, G. 1996, Mon. Not. Roy. Astron. Soc., 281, 1297

\bibitem[{Tegmark {et~al.}(2003)}]{Tegmark:2003ve}
Tegmark, M., {et~al.} 2003, Phys. Rev., D68, 123523

\end{thebibliography}
%\bibliographystyle{apj}

\end{document}